\documentclass[preprintnumbers,amsmath,amssymb,prl]{revtex4}
\usepackage{graphicx}
\begin{document}
\preprint{}
\title{Turning the Cosmological Constant into Black Holes}
\author{Andr\'es Gomberoff}
\author{Marc Henneaux}
\altaffiliation[{Permanent address: }]{Physique Th\'eorique et Math\'ematique,Universit\'e Libre de
Bruxelles,Campus Plaine C.P. 231, B--1050 Bruxelles, Belgium.}
\author{Claudio Teitelboim}
\affiliation{Centro de Estudios Cient\'{\i}ficos (CECS), Valdivia, Chile.}

\date{November 4, 2001}

\begin{abstract}
It is known that  there is  a quantum mechanical tunneling process  which, through nucleation of a membrane, induces
a transition between two de Sitter spaces,  lowering  the cosmological constant.
It is shown in this paper that a different, new,  membrane nucleation  process exists, which, in addition
to lowering the cosmological constant, leaves a black hole behind.  Once a black hole is present,  the  relaxation of the
cosmological constant may proceed via an analog of the old process, which decreases the black hole horizon area, or via the
new process, which increases it.
\end{abstract}
\maketitle
%\pacs{}

%\narrowtext

Microscopic physics would appear to predict an enormously large value for the cosmological constant $\Lambda$ \cite{Weinberg}. However,
the observed value appears to be  very small and positive \cite{obs}. This makes it very appealing to consider $\Lambda$ as a variable that can
evolve dynamically, and relax through quantum mechanical tunneling.

The simplest way of implementing this view is, in four spacetime dimensions,  through the coupling to the gravitational field of a
3--form gauge potential \cite{Aurilia, Duff, Freund, Hawking, HennCT}. In addition to the gauge
potential it is natural to bring in  also its
source, which is a two dimensional membrane that sweeps a three dimensional history.  There exists then a
quantum mechanical process -- analogous to pair creation by an electric field  in two spacetime dimensions -- which
nucleates a membrane and induces a transition between two de Sitter spacetimes, the final one having a  value of the
cosmological constant lower than the initial one \cite{BrownCT, BP,Feng}.

In the case of pair production, the energy of the pair is provided by the decrease in
the energy of the electric field in the space
between the pairs.  Much in the same way, the energy of the membrane which is nucleated may be accounted for,
in physical terms,  by considering the reduction of the cosmological constant inside the bubble as a lowering of
the vacuum energy.

Now, with the dynamical cosmological constant, one has a vast energy sea at one's disposal. It is then natural
to ask whether one
could tap this  reservoir to get from it more than just an expanding membrane, which escapes to infinity, and
whose only effect is to lower the
cosmological constant . Could one perhaps transmute the cosmological background energy into ``localized
matter'' ?. And what localized matter can be more natural  in the context of gravitation theory than a black hole ?.
The purpose of this note  is to answer this question in the affirmative: One may also
nucleate a membrane which contracts to form a black hole, leaving outside of the hole a lower cosmological constant.

To make the point in the most economical manner we will consider  membrane nucleation around a
pre--existent black hole of mass $M_+$ and will then study, in particular, the case $M_+ \rightarrow 0$. This is useful because $M_+ \ne 0$
is the generic situation and the limit $M_+ \rightarrow 0$ is somewhat singular, so that some important features become blurred. The case $M_+ \ne 0$
was considered in \cite{CTplett} in the context of black hole thermodynamics. However,  the new process presented
in this paper, as well as  some important features of the analog of the old process treated  in \cite{BrownCT}, were not discussed there.

We consider an (Euclidean) spacetime element of the form
\begin{equation}
ds^2= f^2(r) dt^2 + f^{-2} dr^2 + r^2 \left( d\theta^2+\sin^2 \theta d\phi^2 \right) \ .
\label{line}
\end{equation}
The antisymmetric field strength tensor takes the form
\begin{equation}
F_{\mu\nu\lambda\rho}=(dA)_{\mu\nu\lambda\rho}=E\sqrt{g}\epsilon_{\mu\nu\lambda\rho} \ .
\label{}
\end{equation}
The history of the  membrane will divide the spacetime in two regions, one which will be called the interior,
labeled by the suffix ``-'' and the other  the exterior, labeled by the suffix ``+''. The boundary in question  may be
described by the parametric
equations
\begin{equation}
r=R(\tau) \ \ \ \ \ \ \ t_{\pm}=T_{\pm}(\tau) \ ,
\label{param}
\end{equation}
where $\tau$ is the proper time, so that its line element reads
\begin{equation}
ds^2= d\tau^2 + R^2(\tau)\left(d\theta^2+\sin^2 \theta d\phi^2 \right) \ ,
\label{memmetric}
\end{equation}
with
\begin{equation}
1=f_{\pm}^2\left(R(\tau)\right) \dot{T}_{\pm}^2 + f_{\pm}^{-2}\left(R(\tau)\right)\dot{R}^2 \ .
\label{propercond}
\end{equation}

In the ``+'' and ``-'' regions the solution of the field equations read
\begin{eqnarray}
f_{\pm}^2 &=& 1-\frac{2M_{\pm}}{r} - \frac{r^2}{l_{\pm}^2} \ ,  \label{Ds} \\
E_{\pm}^2 &=& \frac{1}{4\pi}\left( \frac{3}{l_{\pm}^2}-\lambda \right) \ .
\end{eqnarray}
The actual cosmological constant $\Lambda=3/l^2$ is thus obtained by adding  $\lambda$ (normally taken to be
negative), coming from ``the rest
of physics" and not subject to change, and the contribution $4\pi E^2$, which is subject to dynamical equations.
The discontinuities in the  functions $f^2$ and $E$   across the membrane  are given by
\begin{eqnarray}
f_-^2\dot{T}_- -   f_+^2\dot{T}_+ &=& \mu R  \ .  \label{deltaf}\\
E_{+}- E_{-} &=& q \label{deltaE} \ .
\label{disc}
\end{eqnarray}

Here $\mu$ and $q$ are the tension and charge on the membrane respectively (the action is given in
(\ref{action}) below).
Eq. (\ref{deltaE}) follows from integrating Gauss' law for the antisymmetric tensor across the membrane,
whereas Eq. (\ref{deltaf})
represents the discontinuity in the extrinsic curvature of the membrane when it is regarded as embedded in
either the ``-" or the ``+"
spaces \cite{Israel}.  In writing these equations the following conventions have been adopted, and will be maintained
from here on: $(i)$ The coordinate $t$ increases anticlockwise around the cosmological horizon, $(ii)$
the variable $\tau$ increases when somebody traveling along the curve leaves the interior on his right side.

Equation (\ref{deltaf}) may also be interpreted as the first integral of the equation of motion for the
membrane, which is thus obtained by
differentiating it with respect to $\tau$ (``equations of motion from field equations''). Hence,
satisfying
(\ref{Ds}-\ref{deltaE}) amounts to solving all the equations of motion and, therefore, finding an extremum of
the action.

We are interested in a solution of the equations of motion of the membrane, which is a closed orbit, that will be
 interpreted as an ``instanton'', leading to a reduction of the cosmological constant accompanied by the production
 of a black hole.

To this end we
examine the ``radii of formation''
of the membrane, namely,  the values of $R$ for which $\dot{R}=0$. This is most efficiently analyzed if
one uses the proper
time condition (\ref{propercond}) to rearrange (\ref{deltaf}) to read
\begin{subequations}
\label{DM}
\begin{eqnarray}
\Delta M &=& \frac{1}{2}\left( \alpha^2 -\mu^2 \right) R^3- \mu R ^2f_+^2 \dot{T}_+ \ ,
\label{DMp} \\
&=& \frac{1}{2}\left( \alpha^2 +\mu^2 \right) R^3- \mu R^2f_-^2 \dot{T}_- \ ,
\label{DMm}
\end{eqnarray}
\end{subequations}
where
\begin{equation}
\alpha^2=\frac{1}{l_+^2} - \frac{1}{l_-^2} \ \ \ \ \ \ \Delta M =  (M_--M_+) \  .
\label{alpha}
\end{equation}

The graph of $\Delta M$ as a function of $R$ when $\dot{R}=0$ is given in Fig. 1. There are two branches which merge
smoothly, corresponding to
taking both signs of the square root of $\dot{T}_{\pm}^2$  in (\ref{propercond}) (much as $x=\pm
\sqrt{1-y^2}$ gives the smooth circle
$x^2+y^2=1$).

\begin{figure}[h]
\centering
\includegraphics[width=10cm]{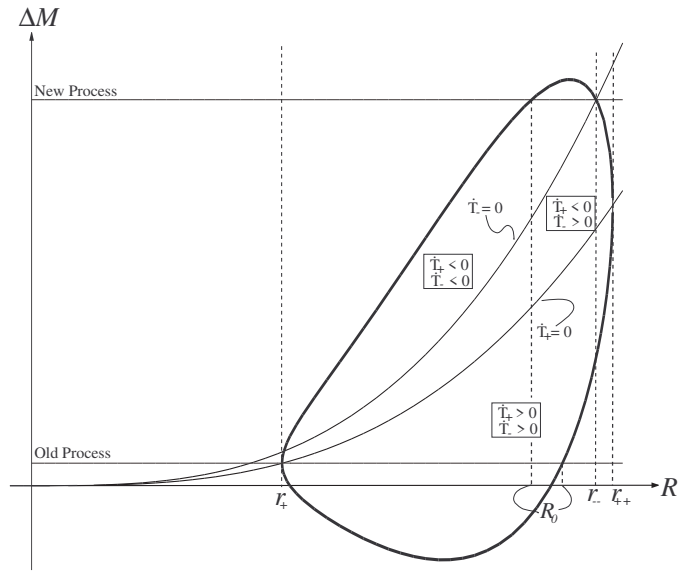}
\caption{Change in the black hole mass as a function of the turning point radius $R$.
The closed curve is the graph of the function $\Delta M(R)$, obtained from  Eq. (\ref{DM}) when $\dot{R}=0$.
If the line $\Delta M=$ constant intersects the curve, it will in general do so at two points.
The corresponding orbits will  bounce between these two radii, but will not close upon themselves. The two special values of
$\Delta M$ for which the orbit closes occur when the second root is $r_+$ (old process), or $r_{--}$ (new process).
Here $r_+$ is the initial black hole horizon and $r_{--}$ is the final cosmological horizon. In each case we call the first root
$R_0$. Also shown in the figure are the curves for which $\dot{T}_{+}$ and $\dot{T}_-$ vanishes, and the three regions
corresponding to different signs for $\dot{T}_{+}$ and $\dot{T}_-$, as inferred from Eq. (\ref{DM}). These regions are crucial
in determining the properties of the instantons (Fig. 2).}
\label{orbits}
\end{figure}

We see that for  each $\Delta M$ within the range shown in fig. 1, there are {\em two} roots, which
correspond to the simultaneous creation of {\em two} membranes,  which have opposite polarities, {\em i.e.}, one has charge $q$,
the other $-q$.
In order for the solution to represent an extremum of the action associated with tunneling, these two roots must be the two turning
points of a single, time symmetric, Euclidean orbit. Now, it may be verified by  studying the orbits which solve (\ref{DM}), that this does not
happen for the generic $\Delta M$. However, there are two special ``mass gaps" for which the orbit indeed closes. They correspond to the limiting cases
when $(i)$ the smaller membrane sits on the horizon of the initial black hole, $(ii)$ the larger membrane sits on the final cosmological
horizon.   In the limiting case, $M_+ = 0$, discussed in \cite{BrownCT}
the smaller membrane has zero radius, disappearing from the problem. However, in the generic case, $M_+\neq 0$, {\em two} membranes are formed
simultaneously. The closed orbits and their properties are discussed in Fig. 2.

The joining of the Euclidean and Lorentzian sections is not devoid of subtleties, and will be discussed  in detail in \cite{GHT}.
The resulting processes  may be described as follows. At a certain moment two membranes are
created and the cosmological constant is lowered in the space in between them.
For the old process, the first membrane materializes
at the radius $R_0$  between the black hole and cosmological horizons, and proceeds to expand, ``collapsing outwards''. At the same time,
the second membrane materializes at a radius $r_+$, the value of $r$ corresponding to the initial black hole horizon.
With  the creation of the membranes, the radius of the black hole horizon decreases ($r_- < r_+$), while that of the cosmological horizon
increases ($r_{--}>r_{++}$).
The second membrane collapses inwards beyond the new black hole horizon. Note that this second membrane can exist only if there is initially a black hole.
The {\em inside} of the first membrane is the ``interior'' in this case.

For the new process the roles of the two horizons are interchanged. The first membrane materializes at the corresponding $R_0$ and proceeds to contract,
collapsing inwards. At the same time, the second membrane materializes right on the final cosmological horizon.
This time the black hole radius increases and the cosmological horizon radius decreases. The second membrane expands beyond the cosmological horizon.
 In this case the first membrane collapses onto the black hole, or forms one if initially there is none.
 Now the {\em outside} of the first membrane is the ``interior''.

Note that with the existence of the new process the symmetry between cosmological and black hole horizons,
which had been lost with the presence of the old process alone, is  restored.  The only remaining asymmetry is
that one can consider the absence of an initial black hole, but one cannot allow, in this context, for the absence of an initial cosmological horizon.

The tunneling processes change both parameters $M$ and $l$ of the Schwarzschild--de Sitter solution. For a
given change  in the cosmological constant, the change in the mass is obtained by setting, in Eq. (\ref{DM}),  $R$ equal to
either the initial black hole horizon, $r_+$ (old process), or
to the final cosmological horizon, $r_{--}$ (new process). This yields
\begin{eqnarray}
\Delta M &=& \frac{1}{2}(\alpha^2 - \mu^2) r_+^3  \hskip 1cm  \mbox{(old process)}\ ,
\label{mass1} \\
\Delta M &=&   \frac{1}{2}(\alpha^2 + \mu^2) r_{--}^3 \hskip 0.8cm  \mbox{(new process)}.
\label{mass2}
\end{eqnarray}
In particular, when $M_+=0$, Eq. (\ref{mass1}) gives $M_-=0$,  in accordance with \cite{BrownCT}.  However, for the
new process, one obtains for $M_-$ the value,
\begin{equation}
M_0= \frac{1}{2}
\frac{   \left(\frac{\Lambda_+}{3}-\frac{\Lambda_-}{3} + \mu^2\right)}{\left(\frac{\Lambda_+}{3}+ \mu^2\right)^{3/2}}
\label{newbh}
\end{equation}
\nopagebreak for the mass of the black hole spontaneously formed out of de Sitter space, when the cosmological
constant decreases from
$\Lambda_+$ to $\Lambda_-$, through nucleation of a pair of  membranes of tension $\mu$.

\begin{figure}[h]
\centering
\includegraphics[width=8cm]{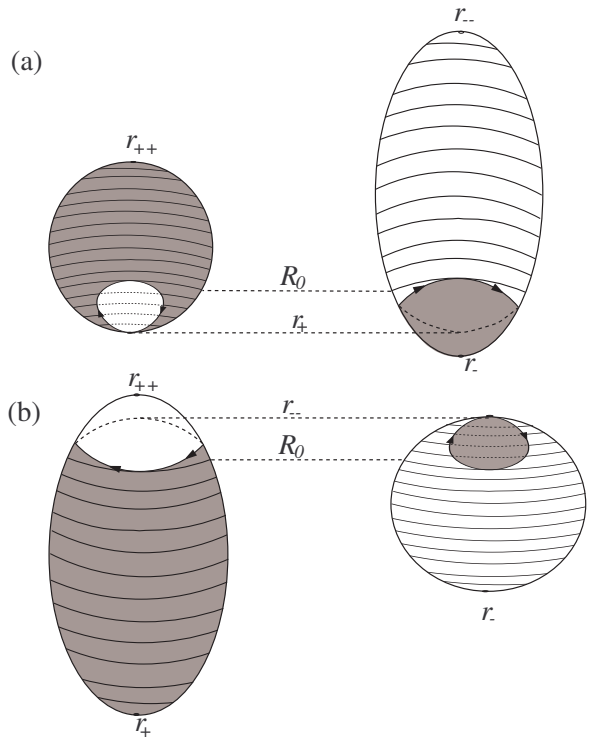}%
\caption{Old and new instantons for tunneling through membrane creation.
The instantons are obtained by glueing, along the history of the membrane,
pieces of two spacetimes with different values of the cosmological constant and the mass.
In both, the old (a) and the new (b) processes, the space on the left corresponds
to the initial value of the parameters, and the one on the right to the final value.
In each case, the white regions in both spaces are deleted  and  the remaining gray regions are joined.
In the old process  the orbit starts at the black hole horizon, $r_+$, of the exterior region,  which is a single point
in $(r,t)$--space. The orbit is therefore closed by construction when seen from the exterior.
 As seen from the interior,  the orbit will close only if one {\em defines} the total amount of time elapsed  $T_-$ as
 being the period. This leads to identifying the endpoints without cutting out a piece of the orbit, and is possible because
 $\dot{T}_- >0$ (Fig. 1). This procedure will, in general, generate a conical singularity at $r_-$.
Also, in order to avoid a conical singularity
at the initial black hole horizon, which is no longer an horizon on the instanton geometry, one must properly adjust
the period of $T_+$ around $r_{++}$. In this way one will generically end up with two conical singularities, at $r_-$
and at $r_{++}$,  (ordinary Euclidean de--Sitter spacetimes has at least one).
The new process  is characterized by the exchange of the roles played by
the "+" and "-" spaces. Here the orbit starts at $r_{--}$, and therefore it is closed by construction as seen by the "-" side.
Now it is necessary to fix the period of $T_{+}$ to close the orbit as seen from the exterior. One ends up again with two
conical singularities.  Note that, in this case, the old cosmological horizon is not in the instanton geometry and therefore there is
no need to worry about a posible conical singularity there, which is just as well because the only adjustable period available,
$T_+$,  has already been used up to close the orbit. One ends up again with two conical singularities, one at $r_+$ the other
at $r_{--}$}
\label{oldinstantonfig}
\end{figure}

 Having established  the existence of the instantons, we pass to discuss  the
probability for each of the two processes.
The general rule is that, in the semiclassical approximation, the probability is the exponential of the
Euclidean action. However, since there are
thermal effects at play due to the presence of event horizons, one must distinguish those from the
underlying quantum mechanical probability for the process.  This is equivalent to subtracting the action of the
background.

When there is only one horizon at play, as it happens for the old process in the absence of an initial black
hole, the identification of the thermal effects can be done neatly since there
exists a  complete  action principle for the coupled system of the gravitational field, the antisymmetric
field, and the membrane. The action,
appropriate for keeping fixed the initial cosmological constant, $\Lambda_+$,  is then equal to
the sum of four contributions. They are
the ``bulk"  gravitational and antisymmetric tensor hamiltonian actions, the membrane action (which includes
couplings with the gravitational and
antisymmetric tensor fields), and  the surface term one fourth of the  cosmological horizon area. The surface term,  is the horizon entropy,
and it stems from the thermal nature of the problem. The rest of the action may be thought of as yielding the
quantum mechanical probability for the process itself. Furthermore, the on--shell value of the bulk Hamiltonian action turns out to be zero, and thus
the action for the background is just one fourth of the area.
Therefore, in the semi--classical approximation,
the quantum mechanical probability for the process is given by  the exponential of the membrane action alone.
\begin{equation}
P = \exp\left[I_{\mbox{\tiny membrane}}\right] \ ,
\label{probability}
\end{equation}
with
\begin{equation}
I_{\mbox{\tiny membrane}}= -\frac{\mu }{4\pi}V_3 + \frac{3}{8\pi}\alpha^2 V_4 \ .
\label{action}
\end{equation}
Here $V_3$ is the $3$--volume of the history of the membrane and $V_4$ is the $4$--volume of its interior.
The $4$--volume term may be
understood as arising from the standard minimal coupling through application of Stoke's theorem. The
coefficient $3\alpha^2/8\pi$ comes from writing
\begin{eqnarray*}
 \frac{3}{8\pi}\alpha^2&=&\frac{3}{8\pi}\left(\frac{1}{l_+^2}-\frac{1}{l_-^2}\right) = \frac{1}{2}\left(E_+^2 - E_-^2 \right)\\
  &=&  \left(E_+-E_-\right)\frac{1}{2}\left(E_+ + E_-\right)= qE_{\mbox{\tiny av}} \ ,
\end{eqnarray*}
where $E_{\mbox{\tiny av}}=(E_+ + E_-)/2$ is the field on the membrane.

When two horizons are present in the instanton, as it happens for the old process  with an initial
black hole or for the new process,
always, there is no complete action principle. This is because one cannot
include appropriate surface terms in the action which ensure that, on--shell, there is no
conical singularity at either horizon (the time periods do not match).
This problem is not a peculiarity of  the membrane, or of
the antisymmetric tensor. It is already
present for the pure gravitational field.
 Thus, already the action for the background is not well defined.
In physical terms this is a reflection of the fact that the black hole and
cosmological horizons are not in thermal equilibrium \cite{GibbHawk}. However, in view of the previous
discussion, it would seem justified to assume that  Eq. (\ref{probability}) remains valid  as the quantum mechanical
probability  for membrane nucleation  for each of the two processes.

Since the main purpose of this communication is to point out that the new process exists, we  will not dwell
here on the actual
evaluation of the semiclassical tunneling probability.  We just point out that it will be necessary to appeal
to approximations to make
the problem  tractable, because in all cases except for the tunneling between two de Sitter spaces discussed
in \cite{BrownCT}, the geometry
has a low degree of symmetry.
For the old process with $M_+=M_-=0$,  the instanton consists  of two  $4$--spheres joined on a
$3$--sphere. However,  in all other cases, for the new and the old processes,  one has two  $S_2\times S_2$'s  joined on
a $S_1\times S_2$. Furthermore, the $S_2$'s are just topological, not metrical spheres.

A fuller account of the results presented in this paper and their possible implications is now in preparation, and will be
published elsewhere \cite{GHT}. However, even at this early stage, it is hard to refrain  from speculating that
the low value of the cosmological constant might point to the presence of mass  in the universe,
in the form of black holes created by its relaxation.

\acknowledgments

We thank Valery Rubakov for an enlightening discussion.
This work  was funded by an institutional grant to CECS of the Millennium Science Initiative, Chile,
and also benefits from the generous support to CECS by Empresas CMPC. AG gratefully  acknowledges
support from FONDECYT grant 1010449  and from Fundaci\'on Andes. AG and
 CT  acknowledge partial support under FONDECYT grant 1010446.
 The work of MH is partially supported by the ``Actions de
Recherche Concert{\'e}es" of the ``Direction de la Recherche
Scientifique - Communaut{\'e} Fran{\c c}aise de Belgique", by
IISN - Belgium (convention 4.4505.86)
and by the European Commission RTN programme
HPRN-CT-00131,  in which he is associated to K. U. Leuven.

\end{document}